*Editorial*

# Investigation of Nanostructures with X-ray Scattering Techniques

**Dominik Kriegner [1,2,]\*, Milan Dopita[3]**

[1] Max Planck Institute for Chemical Physics of Solids, Nöthnitzer Str 40, Dresden, Germany
[2] Institute of Physics ASCR, v.v.i., Cukrovarnicka 10, 162 53, Praha 6, Czech Republic
[3] Department of Condensed Matter Physics, Faculty of Mathematics and Physics of Charles University, Prague, Czech Republic

**\*** Correspondence: dominik.kriegner@gmail.com

The structural investigations of nanomaterials motivated by their large variety and diverse set of applications have attracted considerable attention. In particular, the ever-improving machinery, both in laboratory and at large scale facilities, together with the methodical improvements available for studying nanostructures ranging from epitaxial nanomaterials, nanocrystalline thin films and coatings, to nanoparticles and colloidal nanocrystals allows us to gain a more detailed understanding of their structural properties. As the structure essentially determines the physical properties of the materials, this advances the possibilities of structural studies and also enables a deeper understanding of the structure to property relationships. In this special issue entitled "Investigation of Nanostructures with X-ray Scattering Techniques" five contributions show the recent progress in various research fields. Contributions cover topics as diverse as neutron scattering on magnetic multilayer films, epitaxial orientation of organic thin films, nanoparticle ordering and chemical composition analysis, and the combination of nanofocused X-ray beams with electrical measurements.

In the following we will briefly discuss the contributed papers in order of their submission date. In the first contribution Roland Resel et al. [1] study the epitaxial order of organic thin films. Epitaxy with organic molecules is a rich field due to the wide variety of available organic molecules and their application in organic electronics [2]. An additionally enriching fact is that by the epitaxial growth often metastable, previously unknown, crystal structures of the molecule are found [3]. In their contribution Resel et al. showed that even in cases where the bulk crystal structure prevails during epitaxial growth, the heteroepitaxial orientation of the organic thin film with respect to a single crystalline substrate is a complicated issue on its own. The authors show that for quinquephenyl (5P) films on Cu(110)(2x1)O single crystals the lattice misfit minimization fails to predict the actually occurring orientation of the molecules on the surface. Instead, it is the initial growth stage where the individual molecules fill with their rod-like shape the grooves present in the surface corrugation of the Cu-oxide surface, which determines the epitaxial orientation.

Ryuji Maruyama et al. [4] present a study on the magnetization reversal behavior of thin polycrystalline Fe/Si multilayers by polarized neutron scattering and grazing incidence small-angle neutron scattering. The study shows how the magnetic properties of such multilayers differ from the bulk behavior. In particular, the authors show that in contrast to bulk the lateral correlation length, i.e. the lateral magnetic domain size, varies only insignificantly during de- and re-magnetization. This behavior is explained by the random anisotropy model applicable to systems were the structural grain size is smaller than the ferromagnetic exchange length. Under such situations the coherent rotation of magnetic domains is the dominating process upon magnetization reversal.

Lert Chayanun et al. [5] present a paper about an existing improvement in measurement techniques of nanoscale devices using synchrotron techniques. In particular, they use a nanofocused X-ray beam to study a single nanowire device by various imaging methods. Imaging with (coherent) X-ray beams focused below 100 nm spot size allows insight into the internal structure of nanostructures with resolution limits far below the beam size [6,7]. In the paper in this special issue the authors show how sensitive detection of currents is possible while continuously scanning an X-ray beam across an electrically contacted nanostructure. Simultaneously the tilts and local strains in the nanowire device are determined, which using that method is possible with high spatial and angular resolution [8]. A similar method was previously used to study electrical defects in planar solar cell devices and has great potential to study in-operando devices were the X-ray beam excitation is used to probe the electric field induced changes in nanoscale devices [9].

While in the previous contribution individual nanoscale objects were studied, the remaining two contributions to the special issue show the power of X-ray scattering when it comes to analysis of statistical properties of nanometer sized objects. Lovro Basioli et al. [10] show how the grazing incidence small angle X-ray scattering method can be applied to study ordering in arrays of nanoparticles. Shape, size, arrangement properties and their respective statistical distributions can be determined [11]. The tricky part in the analysis is the development/selection of the correct structural model for the particular system under investigation. As the authors point out a good agreement between experiment and an incorrect model can be obtained yielding wrong values of the structural parameters. The contribution outlines how the analysis of GISAXS patterns can be performed using analytical models and promotes the use of the software package GisaxStudio [12] enabling a straightforward analysis also applicable for less experienced researchers.

In the last contribution to the special issue Jana Šmilauerova et al. [13] investigated the chemical composition of nanoparticles formed inside single crystalline Ti(Mo) alloys. Since the formation of these nanoparticles influences the mechanical properties of the important engineering Ti alloys a better understanding of the nature of the occurring transformation is needed. It was pointed out that contrary previous belief the formation of the nanoparticles is governed by a mixed -mode transformation were structural as well as compositional instabilities play a role [14]. The paper in this special issue determines the chemical composition profiles using a minute analysis of anomalous diffraction data. The employed method is crystallographic phase and chemical selective since diffraction and the proximity of an absorption edge are exploited. The analysis reveals the core/shell nature of the chemical composition upon the formation of $\omega$- phase nanoparticles in the Ti(Mo) alloy and shows how it depends on the processing temperature used during the aging of the alloy.

The diverse set of contributions to this special issue shows that X-ray scattering methods push the limits in the understanding of materials in various fields. Structural research presented here occurs at the forefront of materials research since an improvement of the understanding of the structure almost always leads to deeper insight into the material properties. As nanostructures are interesting for their various and tunable properties, including their magnetic, optic, and thermal properties, the structural studies are pushing the limits in our understanding of matter in general. As these properties can differ greatly in nanoscale objects an improved set of tools/methods are needed to study them. We hope that this special issue could show the state of the art in the field and encourages future work to facilitate a deeper understanding of matter on the nanoscale.

**Acknowledgments:** The Guest Editors thank all the authors contributing in this Special Issue and the Editorial staff of Crystals for their support.

**Funding:** DK acknowledges the support by Operational Programme Research, Development and Education financed by European Structural and Investment Funds and the Czech Ministry of Education, Youth and Sports (Project No. CZ.02.2.69/0.0/0.0/16_027/0008215). The work was further supported by the Ministry of Education of the Czech Republic Grant No. LM2018110 and LNSM-LNSpin. M.D. acknowledges the financial support from the project NanoCent - Nanomaterials Centre for Advanced Applications, Project No. CZ.02.1.01/0.0/0.0/15_003/0000485, financed by ERDF

**Conflicts of Interest:** The authors declare no conflict of interest.